\documentstyle[prl,aps,psfig,twocolumn]{revtex}

\begin{document}
\draft 

\wideabs{
\title{\bf Thermodynamic limit from small lattices of coupled maps}

\author{
R.~Carretero-Gonz\'alez$^{1,}$\cite{rcg:email},
S.~{\O}rstavik$^{1}$, J.~Huke$^{2}$, D.S.~Broomhead$^{2}$
and J.~Stark$^{1}$}
\address{$^{1}$Centre for Nonlinear Dynamics and its
         Applications\cite{CNDA:web}, 
         University College London, London WC1E 6BT, U.K.}
\address{$^{2}$Department of Mathematics, University of Manchester
         Institute of Science \& Technology,
         Manchester M60 1QD, U.K.}

\date{Submitted to Phys. Rev. Lett, March 1999}
\maketitle

\begin{abstract}
We compare the behaviour of a small truncated coupled map lattice with 
random inputs at the boundaries with that of a large deterministic 
lattice essentially at the thermodynamic limit. We find exponential 
convergence for the probability density, predictability, power spectrum, 
and two-point correlation with increasing truncated lattice size. This 
suggests that spatio-temporal embedding techniques using local 
observations cannot detect the presence of spatial extent in such systems 
and hence they may equally well be modelled by a local low 
dimensional stochastically driven system.
\end{abstract}

\pacs{PACS numbers: 05.45.Ra, 05.45.Jn, 05.45.Tp}
}

Observation plays a fundamental role throughout all of physics. Until this
century, it was generally believed that if one could make sufficiently accurate
measurements of a classical system, then one could predict its future evolution
for all time. However, the discovery of chaotic behaviour over the last 100
years has led to the realisation that this was impractical and that there are
fundamental limits to what one can deduce from finite amounts of observed data.
One aspect of this is that high dimensional deterministic systems may in many
circumstances be indistinguishable from stochastic ones. In other words, 
if we have a physical process whose evolution is governed by a large number 
of variables, whose precise interactions are a priori unknown, then we may 
be unable to decide on the basis of observed data whether the system is 
fundamentally deterministic or not. This has led to an informal 
classification of dynamical systems into two categories: low dimensional 
deterministic systems and all the rest. In the case of the former, 
techniques developed over the last two decades allow the characterisation 
of the underlying dynamics from observed time series via quantities such 
as fractal dimensions, entropies and Lyapunov spectra\cite{Schreiber:book}. 
Furthermore, it is possible to predict and manipulate such time series 
in highly effective ways with no prior knowledge of the physical system generating 
the data. In the case of high dimensional and/or stochastic systems, on the 
other hand, relatively little is known about what information can be 
extracted from observed data, and this topic is currently the 
subject of intense research.

Many high dimensional systems have a spatial extent and can best be viewed
as a collection of subsystems at different spatial locations coupled
together.
The main aim of this letter is to
demonstrate that using data observed from a limited spatial region
we may be unable to distinguish such
an extended spatio-temporal system from a local low dimensional system
driven by noise. Since the latter is much simpler, it may in many cases
provide a preferable model of the observed data. On one hand this suggests
that efforts to reconstruct by time delay embedding
the spatio-temporal dynamics of extended
systems may be misplaced, and we should instead focus on developing methods
to locally embed observed data. A preliminary framework for this is
described in \cite{Sakse:98}. On the other hand, these results
may help to explain why time delay reconstruction methods sometimes work
surprisingly well on data generated by high dimensional spatio-temporal
systems, where a priori they ought to fail: in effect such methods only see
a ``noisy'' local system, and providing a reasonably low ``noise level''
can still perform adequately. Overall we see that we add a
third category to the above informal classification: namely that of low
dimensional systems driven by noise and we need to adapt our
reconstruction approach to take account of this. 

We present our results in the context of coupled map lattices (CML's) which
are a popular and convenient paradigm for studying spatio-temporal
behaviour \cite{Kan-Kapral}. In particular, consider a one-dimensional 
array of diffusively coupled logistic maps:
\begin{equation}\label{diffu}
x_i^{t+1} = (1-\varepsilon) f(x_i^t) + {\varepsilon\over 2}(f(x_{i-1}^{t})
+ f(x_{i+1}^{t})),
\end{equation}
where $x_i^t$ denotes the discrete time dynamics at discrete locations
$i=1,\dots,L$, $\varepsilon\in[0,1]$ is the coupling strength and the local 
map $f$ is the fully chaotic logistic map $f(x)=4x(1-x)$. Recent research has 
focused on the thermodynamic limit, $L\rightarrow \infty$, of such dynamical 
systems \cite{Pikovsky:94}. Many interesting phenomena arise in this limit, 
including the rescaling of the Lyapunov spectrum \cite{rcg:sublya} and the 
linear increase in Lyapunov dimension \cite{Parekh:98}.
The physical interpretation of such phenomena 
is that a long array of coupled systems may be thought of as a concatenation 
of small-size sub-systems that evolve almost independently from each 
other\cite{Ruelle-Kan}. As a consequence, the limiting behaviour of an 
infinite lattice is extremely well approximated by finite lattices of 
quite modest size. In our numerical work, we thus approximate the 
thermodynamic limit by a lattice of size $L=100$ with periodic boundary 
conditions.

Numerical evidence\cite{Sakse:98} suggests that the attractor in such a 
system is high-dimensional (Lyapunov dimension approximately 70). If 
working with observed data it is clearly not feasible to use an embedding 
dimension of that order of magnitude. On the other hand, it is 
possible\cite{Sakse:98} to make quite reasonable predictions of the 
evolution of a site using embedding dimensions as small as 4. This 
suggests that a significant part of the dynamics is concentrated in 
only a few degrees of freedom and that a low dimensional model may 
prove to be a good approximation of the dynamics at a single 
site. In order to investigate this we introduce the following truncated 
lattice. Let us take $N$ sites ($i=1,\dots,N$) coupled as in equation 
(\ref{diffu}) and consider the dynamics at the boundaries $x_0^t$ and 
$x_{N+1}^t$ to be produced by two independent driving inputs. The driving 
input is chosen to be white noise uniformly distributed in the interval 
$[0,1]$. We are interested in comparing the dynamics of the truncated 
lattice to the thermodynamic limit case. 

\begin{figure}
\centerline{\psfig{figure=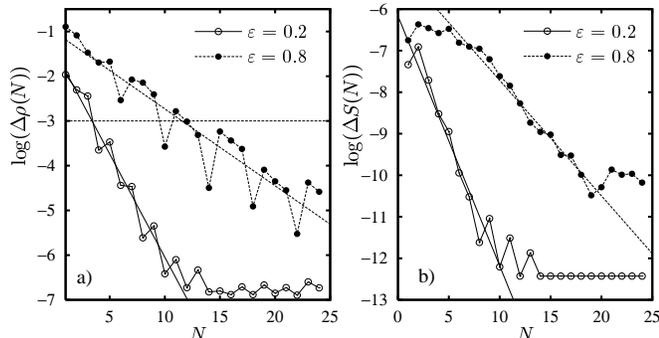,width=8.5cm,silent=,angle=0}}
\caption{Distance between (a) the probability density and (b) the
power spectra in the thermodynamic limit and its truncated lattice
counterpart as the number of sites $N$ in the latter is increased.
}
\label{dist-dens-specG.ps}
\end{figure}

\begin{figure}
\centerline{\psfig{figure=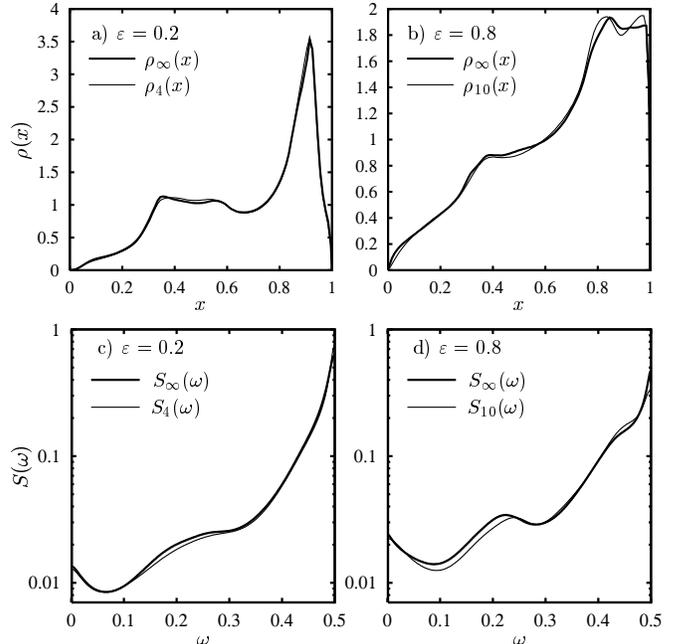,width=8.5cm,silent=,angle=0}}
\caption{Approximating (a,b) the probability density and 
(c,d) the power spectra of the thermodynamic limit
(thick lines) using a truncated lattice (thin lines).
}
\label{dens-specG.ps}
\end{figure}

We begin the comparison between the two lattices by examining
their respective invariant probability density at the central site
(if the number of sites is even, either of the two central
sites is equivalent). For a semi-analytic treatment of the probability
density of large arrays of coupled logistic maps see Lema\^{\i}tre
{\em et al.}\cite{Lemaitre:97}.
Let us denote by $\rho_\infty(x)$ the single site probability density
in the thermodynamic limit and $\rho_N(x)$
the central site probability density of the truncated
lattice of size $N$. We compare the two densities in the ${\cal L}_1$
norm by computing
\begin{equation}\label{dr(N)}
\Delta\rho(N) = \int_0^1{|\rho_\infty(x)-\rho_N(x)|\,dx}
\end{equation}
for increasing $N$. The results are summarised in figure
\ref{dist-dens-specG.ps}.a where $\log(\Delta\rho(N))$ is plotted
for increasing $N$ for different values of the coupling.
The figure suggests that the difference between the densities
decays exponentially as $N$ is increased (see straight lines
for guidance). Similar results were obtained for intermediate 
values of the coupling parameter. 
The densities used to obtain the plots in figure \ref{dist-dens-specG.ps}.a
were estimated by a box counting algorithm by using 100 boxes and $10^8$
points ($10^2$ different orbits with $10^6$ iterations each).
The maximum resolution typically achieved by using these values turns
to be around $\Delta\rho(N) \simeq \exp(-6.5)\simeq 0.0015$. This
explains the saturation of the distance corresponding to
$\varepsilon=0.2$. For $\varepsilon=0.8$ the saturation would
occur for approximately $N=30,35$ given enough computing power
(more refined boxes and more iterations). Nonetheless,
densities separated by a distance of approximately $\exp(-3)\simeq 0.05$
(see horizontal threshold in figure \ref{dist-dens-specG.ps}.a), or less,
capture almost all the structure. Therefore, one recovers the essence
of the thermodynamic limit probability density with a reasonable small
truncated lattice (see figures \ref{dens-specG.ps}.a,b).

Next we compare temporal correlations in the truncated lattice with those
in the full system. Denote by $S_\infty(\omega)$ the power spectrum
of the thermodynamic limit and $S_N(\omega)$ its counterpart for
the truncated lattice.
Figure \ref{dist-dens-specG.ps}.b shows the difference $\Delta S(T)$
in the ${\cal L}_1$ norm between the power spectra of the truncated
lattice and of the thermodynamic limit for $\varepsilon=0.2$ and 0.8
(similar results were obtained for intermediate values of $\varepsilon$).
As for the probability density, the power spectra appear to
converge exponentially with the truncated lattice size.
Note that for large $N$, particularly for small $\varepsilon$,
the difference tends to saturate around $\exp(-12)\approx 10^{-6}$, this
is because the accuracy of our power spectra computations reaches its
limit (with more iterations one can reduce the effects of the saturation).
Our results were obtained by averaging $10^6$ spectra ($|{\rm DFT}|^2$) of 
1024 points each. In figures \ref{dens-specG.ps}.c,d we depict the 
comparison between the spectra corresponding to the thermodynamic limit 
and to the truncated lattice. As can be observed from the figure, the 
spectra for the truncated lattice give a good approximation
to the thermodynamic limit. It is worth mentioning that the spectra
depicted in figures \ref{dens-specG.ps}.c,d are plotted in logarithmic
scale so to artificially enhance the discrepancy of the distance between the
thermodynamic limit and the truncated lattice. The distance corresponding
to these plots lies well below $\Delta S(T)<\exp(-7.5)\approx 5\times 10^{-4}$.
The convergence of the power spectrum is much faster than the one for the
probability density (compare both scales in figures \ref{dist-dens-specG.ps}).

\begin{figure}
\centerline{\psfig{figure=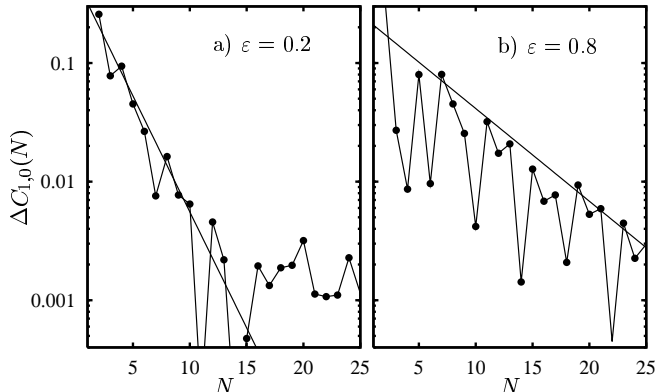,width=8.5cm,silent=,angle=0}}
\caption{Difference of the two-point correlation between the truncated
lattice and the thermodynamic limit for two neighbours at the
same iteration ($C(\xi=1,\tau=0)$).
}
\label{dist-correlG.ps}
\end{figure}

To complete the comparison picture we compute the two-point correlation
\cite{Schreiber:90}
\begin{equation}\label{correl}
C(\xi,\tau) = {\langle uv \rangle - \langle u \rangle \langle v \rangle
\over \langle u^2 \rangle - \langle u \rangle^2},
\end{equation}
where $u = x_i^t$ and $v = x_{i+\xi}^{t+\tau}$. Thus, $C(\xi,\tau)$ corresponds
to the correlation of two points in the lattice dynamics separated by $\xi$
sites and $\tau$ time steps. To obtain the two-point correlation for the
truncated lattice we consider the two points closest to the central site
separated by $\xi$. We then compute $\Delta C_{\xi,\tau}(N)$ defined as
the absolute value of the difference
of the correlation in the thermodynamic limit with that obtained using
the truncated lattice of size $N$. In figure \ref{dist-correlG.ps} we
plot $\Delta C_{1,0}(N)$ as a function of $N$ for $\varepsilon=0.2$ and 0.8.
For $\varepsilon=0.2$, due to limited accuracy of our calculations,
the saturation is reached around $N=10$. Nonetheless it is
possible to observe an exponential decrease (straight lines in
the linear-log plot) before the saturation.
For larger values of $\varepsilon$ the exponential convergence is more
evident (see figure \ref{dist-correlG.ps}.b). Similar results were obtained
for intermediate $\varepsilon$-values.
Note that because the correlation oscillates,
it is not possible to have a point by point exponential decay for
$\Delta C_{1,0}(N)$, however, the upper envelope clearly
follows an exponential decay (see straight lines for guidance).
Similar results were obtained for different values of $(\xi,\tau)$.

The above comparisons were carried out by using the data produced by the
known system (\ref{diffu}). Often, in practice, one is deprived of the
evolution laws of the system. In such cases, the only way to analyse the 
system is by using time series reconstruction techniques. This is particularly
appropriate when dealing with real spatio-temporal systems where, typically,
only a fraction of the set of variables can be measured or when the dynamics 
is only indirectly observed by means of a scalar measurement function.
In the following we suppose that the only available data
is provided by the time series of a set of variables in a small
spatial region. We would like to study the effects on predictability
when using a truncated lattice instead of the thermodynamic limit.

\begin{figure}
\centerline{\psfig{figure=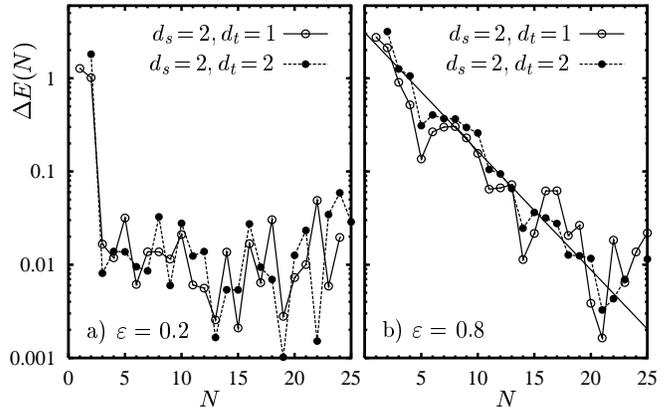,width=8.5cm,silent=,angle=0}}
\caption{Normalised one-step prediction error difference (\ref{error})
between a truncated lattice and the thermodynamic limit for two
spatio-temporal embeddings ($(d_s,d_t)=(2,1)$ and $(d_s,d_t)=(2,2)$)
and different couplings strengths. 
}
\label{predicG.ps}
\end{figure}

Instead of limiting ourselves to one-dimensional time-series (temporal
embedding) we use a mix of temporal and spatial delay embeddings
(spatio-temporal embedding)\cite{Sakse:98}.
Therefore we use the delay map
\begin{equation}\label{delay}
\bbox{X}_i^t = \left(\bbox{y}_i^t,
\bbox{y}_{i-1}^t,\dots,\bbox{y}_{i-(d_s-1)}^t\right),
\end{equation}
whose entries $\bbox{y}_i^t=(x_i^t,x_i^{t-1},\dots,x_i^{t-(d_t-1)})$ are 
time-delay vectors and where $d_s$ and $d_t$ denote the spatial and temporal 
embedding dimensions. The overall embedding dimension is $d=d_s d_t$.
The delay map (\ref{delay}) is used to predict $x_i^{t+1}$.
Note that we are using spatial coordinates only from the
left of $x_i^{t+1}$ ({i.e.}\ $x_j^t$ such that $j\leq i$). An obvious
choice of spatio-temporal delay would be a symmetric
one such as $\bbox{X}_i^t =(x_{i-1}^{t},x_i^t,x_{i+1}^{t})$. However, this
would give artificially good results (for both the full and truncated
lattices) since $x_i^{t+1}$ depends only on these variables
({cf.}\ (\ref{diffu})). This is an artefact of the choice of coupling and
observable and could not be expected to hold in general. Therefore, we use
the delay map (\ref{delay}) in order to ``hide'' some dynamical information
affecting the future state and hence make the prediction problem a
non-trivial one.
The best one-step predictions\cite{Sakse:98} using the delay 
map (\ref{delay}) are typically obtained for $d_s=d_t=2$.
Here we use the two cases $(d_s,d_t)=(2,1)$ and $(d_s,d_t)=(2,2)$;
almost identical results are obtained for higher dimensional
embeddings ($(d_s,d_t)\in[1,4]^2$). Denote by $E(N)$ the normalised
root-mean square error for the one step prediction using the delay map 
(\ref{delay}) at the central portion of the truncated lattice of size 
$N$. The comparison between $E(N)$ and
$E(N\rightarrow \infty)$ is shown in figure \ref{predicG.ps} where
we plot the absolute value of the normalised error difference
\begin{equation}\label{error}
\Delta E(N) = \left|(E(N)-E(\infty))/E(\infty)\right|
\end{equation}
for increasing $N$ and for different spatio-temporal embeddings and
coupling strengths. The figure shows a rapid decay of the prediction
error difference for small $N$ and then a saturation region
where the limited accuracy of our computation hinders any further
decay. For $\varepsilon=0.2$ the drop to the saturation region
is almost immediate while for the large coupling value
$\varepsilon=0.8$ the decay is slow enough to observe an apparently
exponential decay (see fitted line corresponding to $d_s=d_t=2$
for $N=1,\dots,20$), thereafter the saturation region is again 
reached. For intermediate values of $\varepsilon$, the
saturation region is reached between $N=5$ and 20 (results not
shown here). Before this saturation it is possible to observe a rapid
(exponential) decrease of the normalised error difference.
This corroborates again the fact that it seems impossible
in practice to differentiate between the dynamics of the relatively
small truncated lattice and the thermodynamic limit.

All the results in this letter where obtained from the simulation of a
truncated lattice with white noise inputs at the boundaries. Other 
kinds of inputs did not change our observations in a qualitative way.
It is worth mentioning that a truncated lattice with random
inputs with the {\em same} probability density as the thermodynamic
limit ($\rho_\infty(x)$) produces approximatively the same exponential 
decays as above with just a downward vertical shift ({i.e.}\ same decay 
but smaller initial difference).

The properties of the thermodynamic limit of a coupled logistic lattice
we considered here (probability densities, power
spectra, two-point correlations and predictability) were approximated
remarkably well (exponentially close) by a truncated lattice with random
inputs. Therefore, 
when observing data from a limited spatial region,
given a finite accuracy in the computations and
a reasonably small truncated lattice size,
it would be impossible to discern any dynamical difference between the 
thermodynamic limit lattice and its truncated counterpart.
The implications from a spatio-temporal systems time series perspective
are quite strong and discouraging: even though in theory one should be able
to reconstruct the dynamics of the {\em whole} attractor of a spatio-temporal
system from a local time series (Takens theorem \cite{Takens:81}),
it appears that due to the limited accuracy (CPU precision, time and memory 
limitations, measurement errors, limited amount of data)
it would be impossible to test for definite high-dimensional determinism
in practice.

The evidence presented here suggests the impossibility of reconstructing the
state of the whole lattice from localised information. It is natural to ask
whether we can do any better by observing the lattice at many (possibly all)
different sites. Whilst in principle this would yield an embedding of the
whole high-dimensional system, it is unlikely to be much more useful in
practice. This is because the resulting embedding space will be extremely 
high dimensional and any attempt to characterise the dynamics, or fit a 
model will suffer from the usual "curse of high dimensionality". In particular, 
with any realistic amount of data, it will be very rare for typical points to 
have close neighbours. Hence, for instance, predictions are unlikely to be 
much better than those obtained from just observing a localised part of the 
lattice.

If one actually wants to predict the behaviour at many or all sites, our 
results suggest that the best approach is to treat the data as coming from 
a number of uncoupled small noisy systems\cite{Stark:stochastic}, 
rather than a single large system. 
Of course, if one has good reason to suppose that the system is spatially 
homogeneous, one should fit the same local model at all spatial 
locations, thereby substantially increasing the amount of available data.

This work was carried out under an EPSRC grant (GR/L42513). JS would
also like to thank the Leverhume Trust for financial support.


\end{document}